FRONT MATTER

**Title: Filaments of Galaxies as a Clue to the Origin of Ultra-High-Energy Cosmic Rays (Short Title: UHECRs and Filaments of Galaxies)**

**Summary (or teaser):** There are filaments of galaxies, connected to the Virgo Cluster, around TA hotspot, and they should provide us a clue for the origin of hotspot UHECRs.

**Authors:** Jihyun Kim[1], Dongsu Ryu[1*], Hyesung Kang[2], Suk Kim[3], Soo-Chang Rey[4]

**Affiliations**
[1] Department of Physics, UNIST, Ulsan 44919, Korea
[2] Department of Earth Sciences, Pusan National University, Busan 46241, Korea
[3] Korea Astronomy & Space Science Institute, Daejeon 34055, Korea
[4] Dept of Astronomy & Space Science, Chungnam National Univ, Daejeon 34134, Korea
[*] Corresponding author. E-mail: ryu@sirius.unist.ac.kr

**Abstract**

Ultra-high-energy cosmic rays (UHECRs) are known to come from outside of our Galaxy, but their origin still remains unknown. The Telescope Array (TA) experiment recently identified a high concentration in the arrival directions of UHECRs with energies above $5.7 \times 10^{19}$eV, called hotspot. We here report the presence of filaments of galaxies, connected to the Virgo Cluster, on the sky around the hotspot, and a statistically significant correlation between hotspot events and the filaments. With 5-year TA data, the maximum significance of binomial statistics for the correlation is estimated to be $6.1\sigma$ at correlation angle $3.4°$. The probability that the above significance appears by chance is $\sim 2.0 \times 10^{-8}$ ($5.6\sigma$). Based on this finding, we suggest a model for the origin of TA hotspot UHECRs; they are produced at sources in the Virgo Cluster, and escape to and propagate along filaments, before they are scattered toward us. This picture requires the filament magnetic fields of strength $\gtrsim 20$nG, which need to be confirmed in future observations.

**MAIN TEXT**

**Introduction**

Since their first discovery in the early 1960's, ultra-high-energy cosmic rays (UHECRs) above the so-called Greisen–Zatsepin–Kuz'min (GZK) energy of $E_{GZK} \approx 5 \times 10^{19}$ eV (1,2) have been observed for more than several decades (3,4). However, their origin still remains unresolved. Super-GZK UHECRs are believed to originate from extragalactic objects, as they cannot be confined by the Galactic magnetic field and accelerated within the Galaxy. On the other hand, a substantial fraction of them should come from objects in the local universe within $\sim 100$ Mpc, since the interaction with cosmic microwave background photons limits their propagation distance. One of the obstacles in the study of super-GZK UHECRs has been a small number of observed events. To overcome it, international collaboration projects, such as the Pierre Auger Observatory and the Telescope Array (TA), which deployed large numbers of ground detectors along with fluorescence stations, were launched to collect UHECR data of high quantity and quality.

The collaborations have investigated anisotropies in the arrival directions of observed UHECR events as well as their correlations with nearby extragalactic objects to search for the sources (*5-8*).

The TA collaboration recently reported a high concentration of UHECR events in the sky, called hotspot, centered at RA = 146.7° and DEC = 43.2°, from a study of intermediate-scale anisotropy in the arrival directions of UHECRs above $5.7 \times 10^{19}$ eV, using the scintillator surface detector (SD) data collected over a 5 year period (*8*). The Li-Ma statistical significance of the hotspot was reported to be 5.1σ, and the probability that the hotspot appears by chance was $3.7 \times 10^{-4}$ (3.4σ). Despite the high statistical significance, however, astronomical objects that could be responsible for UHECRs have not been identified around the hotspot, leaving its nature mysterious.

In this report, we describe a new finding that there are filaments of galaxies connected to the Virgo Cluster, around the hotspot; their locations in the sky are correlated to hotspot events with a high statistical significance. Based on this finding, we suggest a model for the origin of the hotspot

## Results

Figure 1 shows the distribution of nearby galaxies within 50 Mpc, taken from the HyperLEDA database (*9*, and see SM for a short description of the database), along with the arrival directions of 72 TA events above $5.7 \times 10^{19}$ eV, used in the hotspot study (*8*). The excess of the TA events within the circle of ~20° radius around the hotspot center is evident, while there is no sign of an obvious excess of nearby galaxies. On the other hand, a deficit in events toward the large concentration of galaxies, the Virgo Cluster, is noticeable. It is the closest galaxy cluster in the direction of RA = 187.7° and DEC = 12.4° at the distance of about 16.5 Mpc, and contains ~1500 galaxies (*10,11*), including M87, one of the brightest radio sources in the sky (*12*). However, due to its position, which is about 50° apart from the hotspot center, the Virgo Cluster has been excluded from consideration for the origin of the TA hotspot so far.

It has been known that there are structures made of galaxies, which are linked to the Virgo Cluster. For instance, a number of elongated structures, called clouds or spurs, were reported in an early study to reveal the Local Supercluster (*13*). More recently, using a dataset taken from HyperLEDA, which in particular includes a large number of low-luminosity, dwarf galaxies, seven filaments of galaxies and one sheet were either reaffirmed or newly identified around the Virgo Cluster. From a Hubble diagram analysis in the Virgo-centric reference frame, it was found that "six" filaments have galaxies infalling toward the Virgo Cluster, so they are dynamically connected to the cluster (*14* and see SM for a further description of the filaments). Figure 2 shows the six filaments, labeled as F1-6. Among the six, "three" filaments, F1-3, are located within ~20° from the hotspot center, and a number of the TA events are found to be aligned along them.

The significance of the association between the filaments F1-3 and the TA events was examined with the following two statistics for correlation. One is the cumulative probability, $P(\theta)$, of binomial statistics to observe excesses of events within the angular distance $\theta$ from the spine of the closest filament. Figure 3 shows $P(\theta)$; it has the minimum value, $1.23 \times 10^{-9}$, at $\theta = 3.4°$, within which $n_{obs} = 18$ are observed among the total $N = 72$ events while $n_{exp} = 3.12$ are expected. It is equivalent to a two-sided Gaussian significance of 6.1σ. At the average distance of the filaments, ~15$h^{-1}$ Mpc, the correlation angle, $\theta = 3.4°$, corresponds to ~1.3 Mpc, which is about the radius of typical filaments. The other is the probability that this significance appears by chance. This post-trial probability, calculated using a Monte-Carlo simulation, is $2.0\times10^{-8}$, equivalent to a significance of $5.6\sigma$. For reference, the statistical significance of the correlation of the TA events with all the six filaments connected to the Virgo Cluster can be inspected, as well. The two statistics give $5.6\sigma$ and $5.1\sigma$, respectively. Lower significances agree with the visual impression that there are not many TA events around the filaments F4-6.

The estimated significance for the correlation between the filaments F1-3 and the TA events is rather high. This could be partially due to the small number of events. The TA events from 7 year SD data were shown in a figure (*15*); from the figure it can be estimated that $n_{obs}$ = 21 events among the total $N$ = 110 events at $\theta$ = 3.4°, giving the binomial statistical significance of $5.7\sigma$. It is still sufficiently large, but a bit smaller than that from the 5 year data. So we point that the statistics quoted here needs to be further updated as more events accumulate.

**Discussion**

The correlation between the TA hotspot events and the filaments F1-3, if it is real, should provide us a clue for the origin of the UHECR events. The straightforward interpretation would be that UHECRs are produced at sources located in the filaments. However, by considering that the members of the filaments are mostly normal, not active, galaxies and many of them are dwarf galaxies (*14*) and also the total number of those galaxies is small, that would be a less likely scenario.

The correlation can also arise, if UHECRs are produced inside the Virgo Cluster and escape preferentially to the filaments, before some of them are scattered toward us, as schematically shown in Figure 4. Galaxy clusters host a number of potential accelerators of UHECRs, listed in the so-called Hillas diagram (*3*). As noted above, there is a powerful radio galaxy, M87, in the Virgo Cluster, and it has been long argued that radio galaxies could be the sources of UHECRs (*16*). A large number, ~1500, of cluster galaxies means that there is a good chance of previous episodes of transient objects like gamma-ray bursts (GRBs), which have been also considered as possible sources (*17*). It was also suggested that cluster-scale shock waves could accelerate UHECRs (*18*). In addition, the Virgo Cluster is known to be young and dynamically active with sub-structures (*10*), suggesting a high possibility of energetic phenomena. However, the detailed processes of UHECR production need to be further investigated.

The medium between galaxies in clusters, the intracluster medium (ICM), is known to be magnetized at the level of the order of $\mu$G strength (*19*). In the Virgo Cluster, for instance, the Faraday rotation measure (RM) of a few hundreds of rad m$^{-2}$ was observed in the M87 jet (*20*). While most of the RM is likely to be associated with the sheath of the jet, it also implies the presence of ICM magnetic fields. In addition, constrained simulations, which were designed to reproduce the density and magnetic field distributions in the Local Universe, supported the magnetic fields of $\mu$G in the Virgo cluster (*21*). Such magnetic fields strongly influence the trajectory of UHECRs. The gyroradius of charged particles is given as

$$r_g \sim 50 \text{ kpc} \left(\frac{E/Z}{5\times10^{19} \text{ eV}}\right)\left(\frac{B}{1\ \mu G}\right)^{-1},$$

where $E$ and $Z$ are the particle's energy and charge. Hence, UHECRs, if produced inside the Virgo Cluster, should be rather tightly confined by the cluster magnetic fields.

It is believed that magnetic fields also exist in the medium of galaxy filaments, although their nature is less well known than that of ICM fields. In a dataset of NVSS (the NRAO

VLA Sky Survey) RMs of extragalactic radio sources, the contribution due to the magnetic fields in the cosmic web was estimated to be $\sigma_{RM} \sim 6$ rad m$^{-2}$ (*22*). It was shown that the dispersion of RM can be reproduced, if filaments are magnetized at the level of ~10 nG (*23*). More recently, a search for the synchrotron emission from the cosmic web set an upper limit of ~30 nG in filaments (*24,25*). Such fields were also predicted in theories and simulations; for instance, the fields could develop through turbulence dynamo during the formation of the large-scale structure (LSS) of the universe (*26*). In addition, in the hierarchical formation of the LSS, matter falls to clusters along filaments, dragging filament magnetic fields into clusters. So field lines are expected to be connected to those of clusters. If such a picture is applied to the Virgo Cluster and the neighboring filaments, UHECRs, which were produced and have roamed around for a while in the cluster, would preferentially escape to the filaments through connected field lines.

Along filaments, whose radius is of the order of Mpc, UHECRs can be confined and guided by the magnetic fields of strength $\gtrsim 20$ nG, if they are protons. In addition to the regular component, the filament magnetic fields should have the random (or turbulent) component, which can scatter UHECRs. The average scattering angle depends on the characteristics of turbulent fields as

$$\delta \sim f \times \frac{\pi}{2} \left(\frac{5 \times 10^{19} \text{ eV}}{E/Z}\right)\left(\frac{L}{25 \text{ Mpc}}\right)^{1/2} \left(\frac{l_c}{1 \text{ Mpc}}\right)^{1/2} \left(\frac{B_{random}}{20 \text{ nG}}\right),$$

where $L$ is the propagation distance and $l_c$ is the coherence length of random magnetic fields, $B_{random}$. Here, $f$ is a reduction factor of the order of unity (*27*). So UHECRs, again if they are protons, could be scattered out of filaments, after they propagate a few tens of Mpc. These estimates are consistent with the picture presented in Figure 4, provided that the filament magnetic fields have the strength of $\gtrsim 20$ nG and $l_c$ of a sizable fraction of the filament radius.

The propagation and scattering of UHECRs in the cosmic web were previously studied through numerical simulations (*28, 29*); it was shown that the arrival direction of super-GZK protons could be substantially deflected. In order to further examine the plausibility of the picture in Figure 4, we followed and inspected the trajectories of UHE protons in the intergalactic space of a simulated model universe (see Materials and Methods below). Figure 5 shows two of them with $6 \times 10^{19}$ eV around a cluster of $T = 3.5$ keV; here, the magnetic field strength is 1.5 $\mu$G in the cluster core, ~0.1 $\mu$G in the cluster outskirts, and a few tens of nG in filaments. Figure 6 shows their displacement from the launching positions, color-coded with the local magnetic field strength, as a function of time. The two UHE protons roam around the cluster core for ~40 Myrs (case 1) and ~100 Myrs (case 2) and move to the outskirts; they escape through filaments, propagate ~10 $h^{-1}$ Mpc (case 1) and ~35 $h^{-1}$ Mpc (case 2), and finally scatter to voids. However, we should note that the specific trajectory depends sensitively on the exact configuration of magnetic fields. So the relevant statistics such as the fraction of the UHE protons confined within the cluster during the GZK time (~300 Myrs), the fraction of the UHE protons escaping to filaments, and the propagation distance along filaments rely on accurate modeling of magnetic fields.

For the true test of the picture, hence, we need to know the magnetic field distribution in the regions of the Virgo Cluster and the hotspot, especially, the field line topology from the cluster to the filaments F1-3. As pointed above, however, the magnetic fields in the cosmic web including those in the filaments F1-3 are not yet well known. But upcoming large-scale astronomical projects, such as the Square Kilometre Array (SKA), list the exploration of intergalactic magnetic fields as one of the key sciences (*30*). So the magnetic fields in the cosmic web may be better constrained in the near future.

A missing piece of the puzzle in this picture is why there is no obvious excess of the TA events toward the filaments F4-6. One possible explanation could be that if cluster magnetic field lines are preferentially connected to those of the filaments F1-3, UHECRs could escape mostly toward them, just like fast solar wind particles flowing out through open magnetic field lines associated with coronal holes of the Sun (*31*). We also mention two interesting observations; first, brightest elliptical galaxies are aligned toward the foot direction of F1-3, hinting that the galaxies might form a structure connected to the filaments, and second, the jet of M87 expands toward that direction as well (see SM for a further description of these observations and *32-34*). In addition, we point a geometrical factor that should contribute to the bias; the region containing the upper filaments, F1-4, is subject to a larger TA exposure than that of the lower filaments, F5-6. In fact, although an event has been observed within $\theta = 3.4°$ of F4, there is no TA event around F5 and F6.

We need to comment on the implication of the Galactic magnetic field. Outside the Galactic disk, it has the strength of ~ $\mu$G and the coherence scale of ~ kpc (*35*). The magnetic field can deflect the trajectory of UHE protons above $5.7\times10^{19}$ eV by a few degree (*36*). The deflection should be much larger if they are heavier nuclei. Hence, the correlation presented in this paper, if real, should indicate that UHECRs are mostly protons.

In this paper, we mainly concerned the origin of the TA events around the hotspot. However, there are super-GZK UHECRs observed in other parts of the sky, which were likely produced at sources outside of the Virgo Cluster. Considering most of them should still come from nearby sources within the GZK-horizon as mentioned above, understanding the cosmic-web structure, including the distribution of galaxy filaments, in the local Universe would be essential to the exploration of the origin of UHECRs. As for the identification of the filaments connected to the Virgo Cluster, however, the task can be completed only when very deep spectroscopic data of low-luminosity, dwarf galaxies are available, so it should be left as a future work.

**Materials and Methods**

The binomial statistics presented above was calculated as follows. Let $n_{obs}$ be the number of events observed within the angular distance $\theta$ from the spine of the closest filament, and $n_{exp}$ be the number expected from the isotropic distribution after the geometrical exposure of the TA experiment, $g(\psi) = \sin\psi\cos\psi$ ($\psi$ is the zenith angle), is taken into account. Then, the cumulative probability to observe $n_{obs}$ or more events within $\theta$ is calculated as

$$P(\theta) = \sum_{x=n_{obs}}^{N} \binom{N}{x} p^x (1-p)^{N-x},$$

where $N$ is the total number of events and $p = n_{exp}/N$. It shows the minimum value, $1.23\times10^{-9}$, at $\theta = 3.4°$, as mentioned above.

The probability that the above significance appears by chance was estimated as follows. Through a Monte-Carlo simulation, $10^9$ data sets were generated; each contains 72 events within the TA field of view, which were drawn randomly after the exposure factor, $g(\psi)$, was taken into account. The cumulative binomial probability as a function of $\theta$ was calculated for each data set, and the cases with the minimum probability smaller than $1.23 \times 10^{-9}$ were counted. This yields the post-trial probability of $2.0 \times 10^{-8}$.

The model universe, where the trajectories of UHE protons were followed, was generated through a numerical simulation for the LSS formation using a particle-mesh/Eulerian cosmological hydrodynamics code (*37*). Assuming a $\Lambda$CDM cosmological model, the following parameters were employed: baryon density $\Omega_{BM} = 0.044$, dark matter density $\Omega_{DM} = 0.236$, cosmological constant $\Omega_{\Lambda} = 0.72$, Hubble parameter $h \equiv H_0/(100 \ \mathrm{kms^{-1}Mpc^{-1}}) = 0.7$, rms density fluctuation $\sigma_8 = 0.82$, and primordial spectral index $n = 0.96$. A cubic box of comoving size of 57 $h^{-1}$ Mpc with periodic boundaries, divided into $1650^3$ uniform grid zones, was employed; the grid resolution is 34.5 $h^{-1}$ kpc, which is smaller than the gyroradius of UHE protons in most zones. Three clusters with X-ray weighted temperature $T \gtrsim 3$ keV formed within the simulation volume, and a cluster with $T = 3.5$ keV was selected as the source cluster of UHE protons (see Figure 5).

Assuming that the intergalactic magnetic fields were seeded by the Biermann battery mechanism, their evolution and amplification were followed (*38*). However, with the numerical resolution employed, the cluster magnetic fields are not amplified to the level of observed strengths (*39*). So the magnetic field strength in the core (within 1 $h^{-1}$ Mpc from the X-ray center) of the source cluster was rescaled to ~1 $\mu$G; then it became ~0.1 $\mu$G in the cluster outskirts, and a few tens of nG in filaments, as mentioned above. At random positions within the cluster core, UHECRs with $6 \times 10^{19}$ eV were injected toward random directions, and their trajectories were followed with the relativistic equation of motions for charged particles under magnetic field.

## Supplementary Materials

Supplementary Materials and Methods
Fig. S1. Leo Minor filament, or F3, in the three-dimensional Supergalactic coordinates.
Fig. S2. Six filaments dynamically connected to the Virgo Cluster, and their names.
Fig. S3. Blow-up of a sky region including the Virgo Cluster and the filaments F1-3.
Data File S1. Sky positons of galaxies for six filaments connected the Virgo Cluster.

**Acknowledgments**

**General**: We thank the Telescope Array (TA) collaboration, especially Prof. H. Sagawa, for discussions.

**Funding:** This work was supported by the National Research Foundation of Korea through grants 2016R1A5A1013277, 2017R1A2A1A05071429, 2017R1D1A1A09000567, 2017R1A5A1070354, and 2018R1A2B2006445.

**Author contributions:** J.K. calculated the statistics for the correlation between hotspot events and filaments, and the trajectories of UHECRs around simulated clusters. D.R. led the project. D.R. and H.K. wrote the manuscript. S.K. and S.-C.R. provided the galaxy data for filaments around the Virgo Cluster. All authors participated in scientific discussions.

**Competing interests:** The authors declare that they have no competing interests.

**Data and materials availability:** The correlation statistics presented in the Results were calculated with the TA data for 72 events above 5.7 $\times 10^{19}$ eV and the galaxy data for filaments around the Virgo Cluster. The TA data was published (*8*) and are downloadable from the journal website, http://iopscience.iop.org/2041-8205/790/2/L21/suppdata/apjl498370t1_mrt.txt. The galaxy data was described before (*14*), and are listed in SM.

**Figures and Tables**

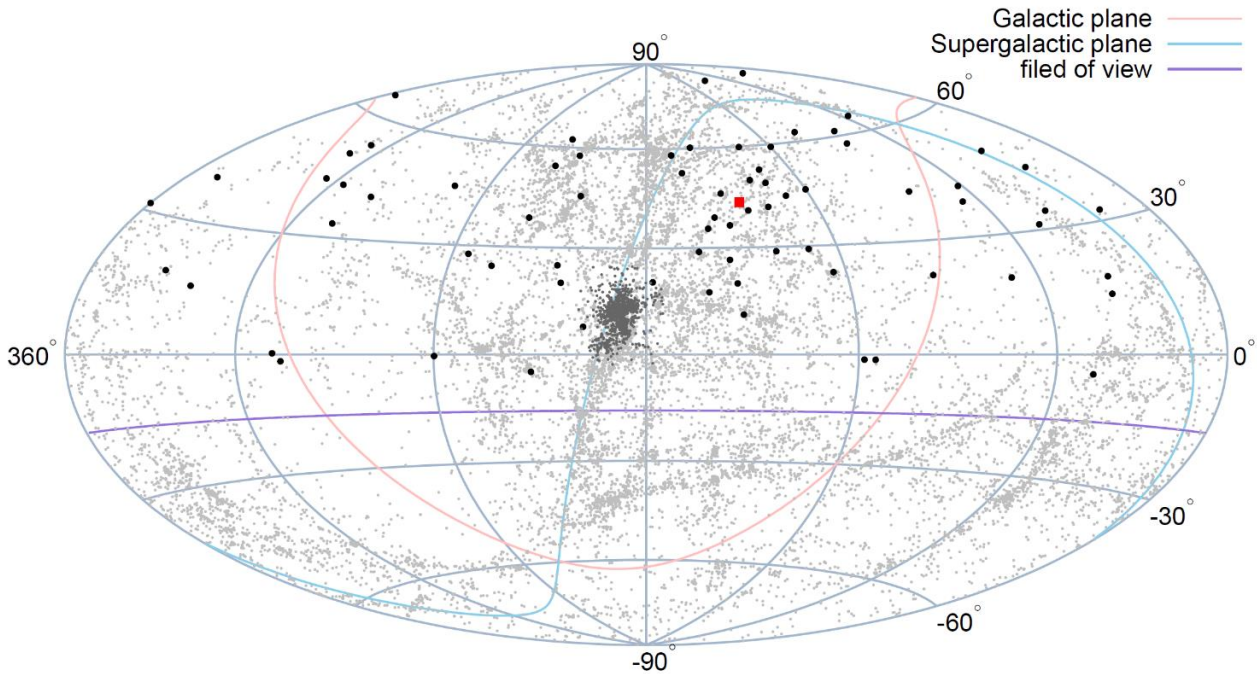

**Figure 1. Hammer projection of nearby galaxies within 50 Mpc (gray dots) and TA UHECR events above 5.7 ×10$^{19}$ eV (large black dots) in equatorial coordinates.** The concentration of galaxies, represented by dark gray dots, is the Virgo Cluster. The red square marks the center of the TA hotspot. The pink and sky-blue lines draw the Galactic and Supergalactic planes, respectively. The purple line indicates the field of view limit of the TA experiment.

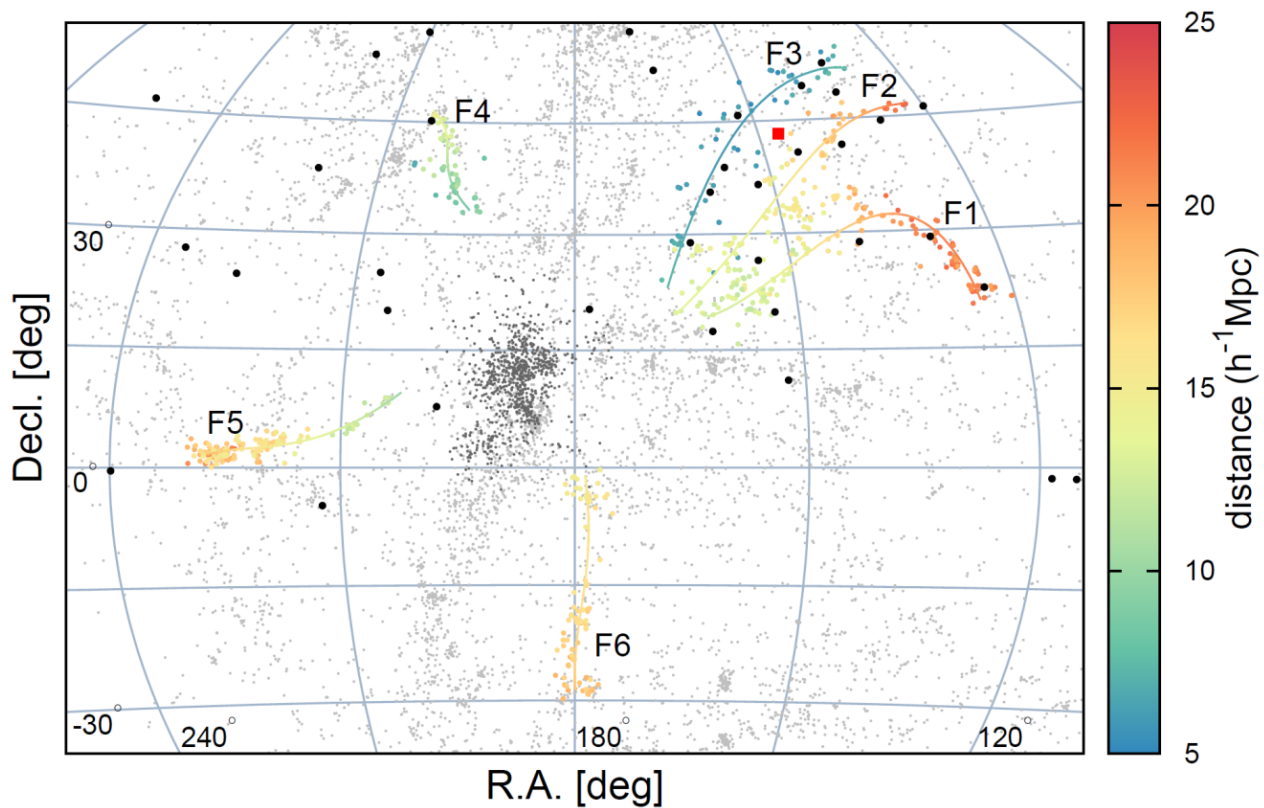

**Figure 2. Six filaments of galaxies connected to the Vigo Cluster, F1-6.** The color dots show the galaxies that belong to the filaments, and the color lines draw the spines of the filaments. The color codes the distance from us to the galaxies and the spines ($h$ is the Hubble parameter). The TA events above 5.7 $\times 10^{19}$ eV (large black dots), nearby galaxies within 50 Mpc (gray dots), the Virgo Cluster galaxies (dark gray dots), and the center of the TA hotspot (red square) are also shown.

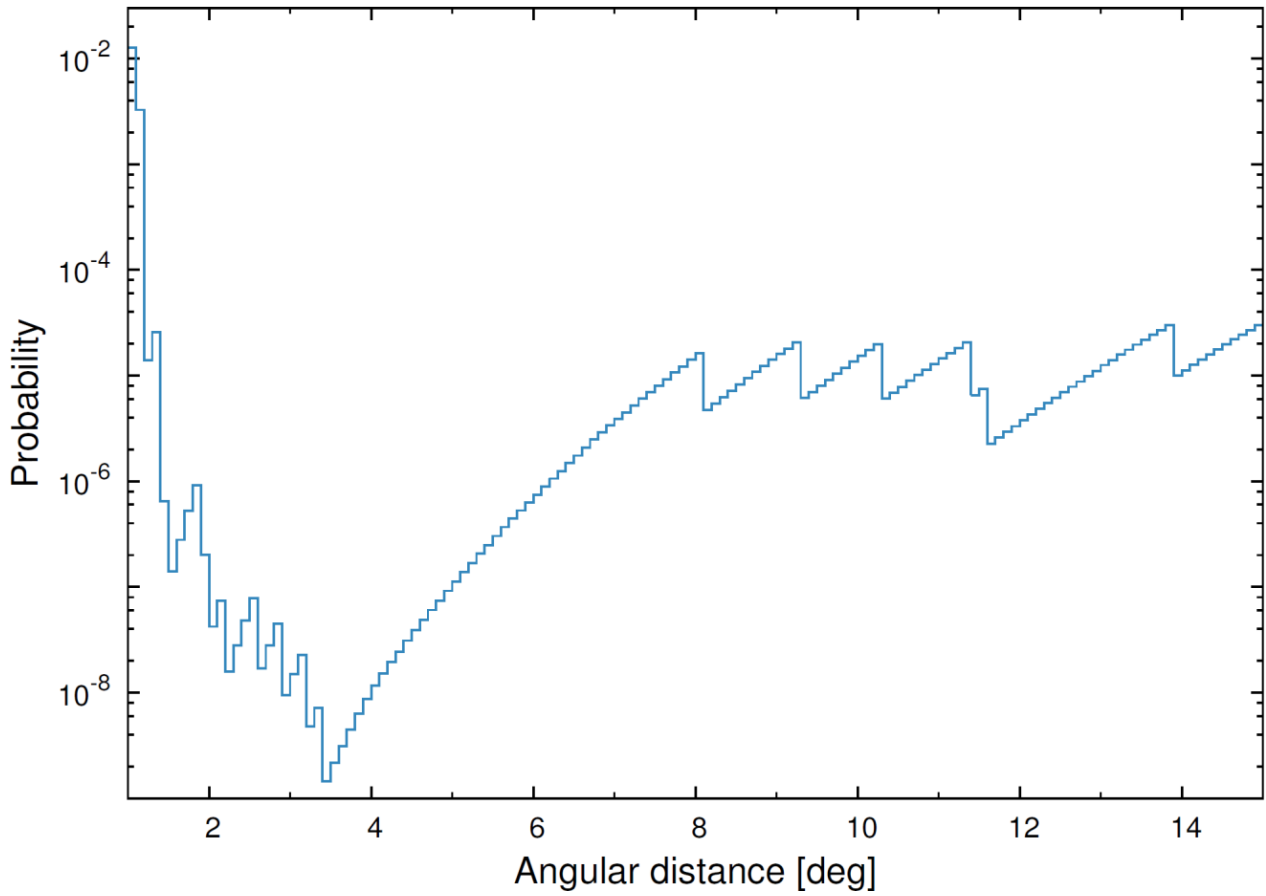

**Figure 3. Cumulative binomial probability** for the excess of TA hotspot events around the filaments F1-3 as a function of angular separation.

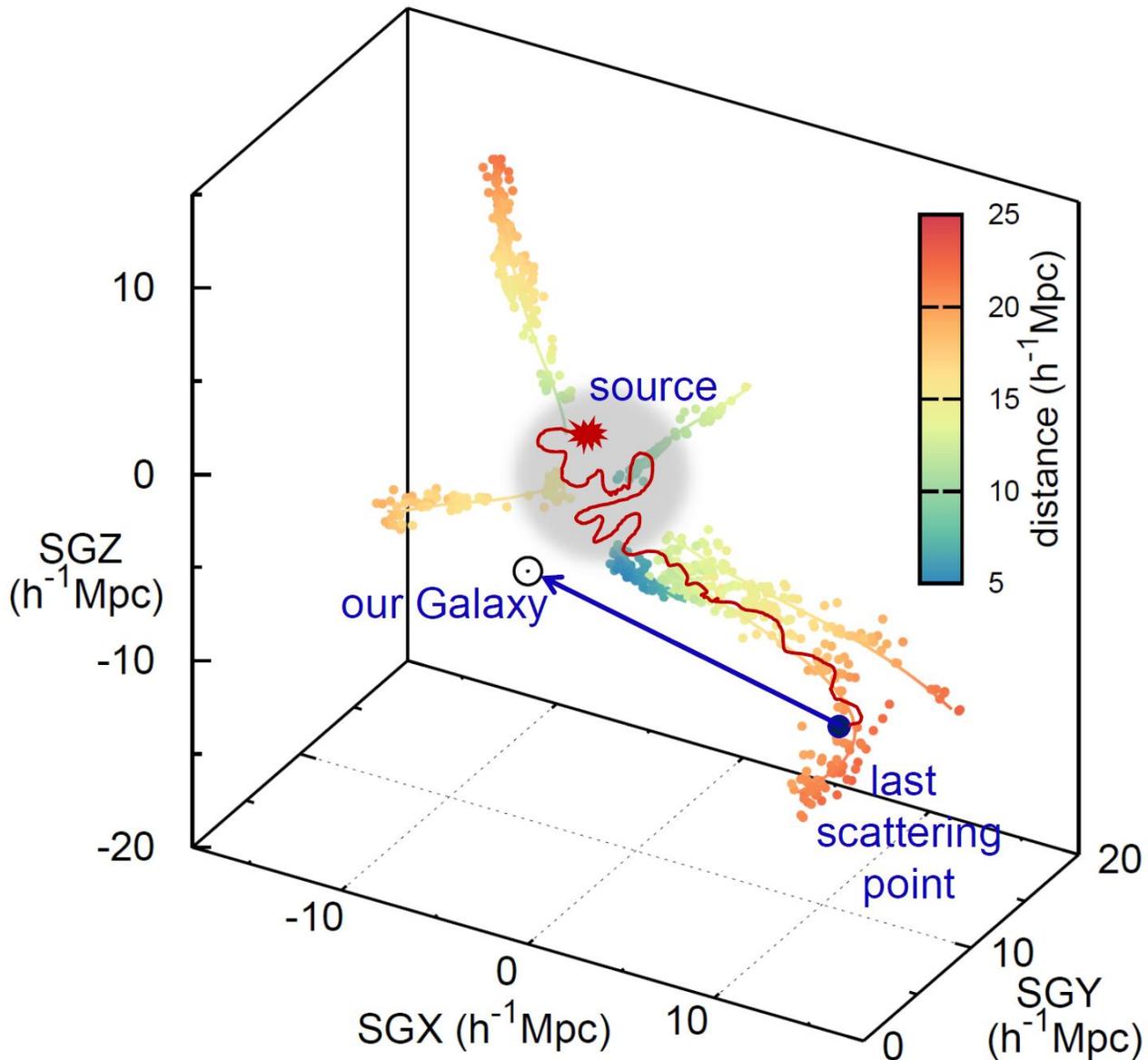

**Figure 4. Schematic drawing of a model for the origin of TA hotspot events.** UHECRs are postulated to be produced at a source or sources inside the Virgo Cluster. After they are confined by cluster magnetic fields and roam around for a while, UHECRs escape to the filaments connected to the cluster. Then, they propagate along the filaments. Some of them are eventually scattered by the random component of magnetic fields, and may come to our Galaxy. Here, the Virgo Cluster, represented by a gray circle, and the filaments F1-6 are plotted in the Supergalactic coordinates. Our Galaxy is located at the coordinate origin.

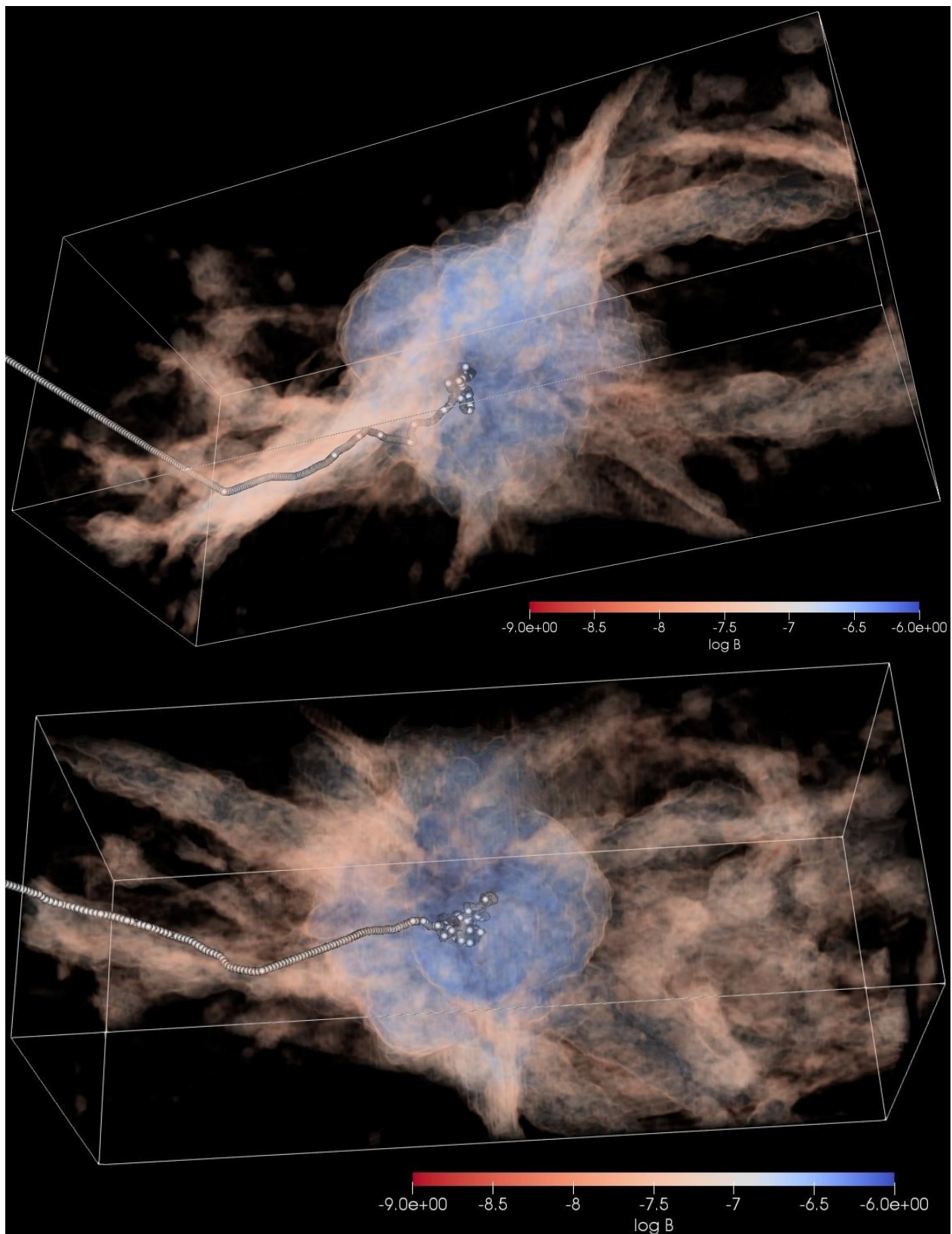

**Figure 5. Trajectories of two UHE protons with $6 \times 10^{19}$ eV around a simulated cluster.** The cluster has the X-ray weighted temperature, $T = 3.5$ keV, and the core magnetic field of ~1.5$\mu$G. Colors code the magnetic field strength; the cluster is represented by the blue tone, while filaments are by the red tone. The box drawn with white lines has the volume of 42×17.5×17.5 ($h^{-1}$ Mpc)$^3$. The trajectories are plotted with white dots.

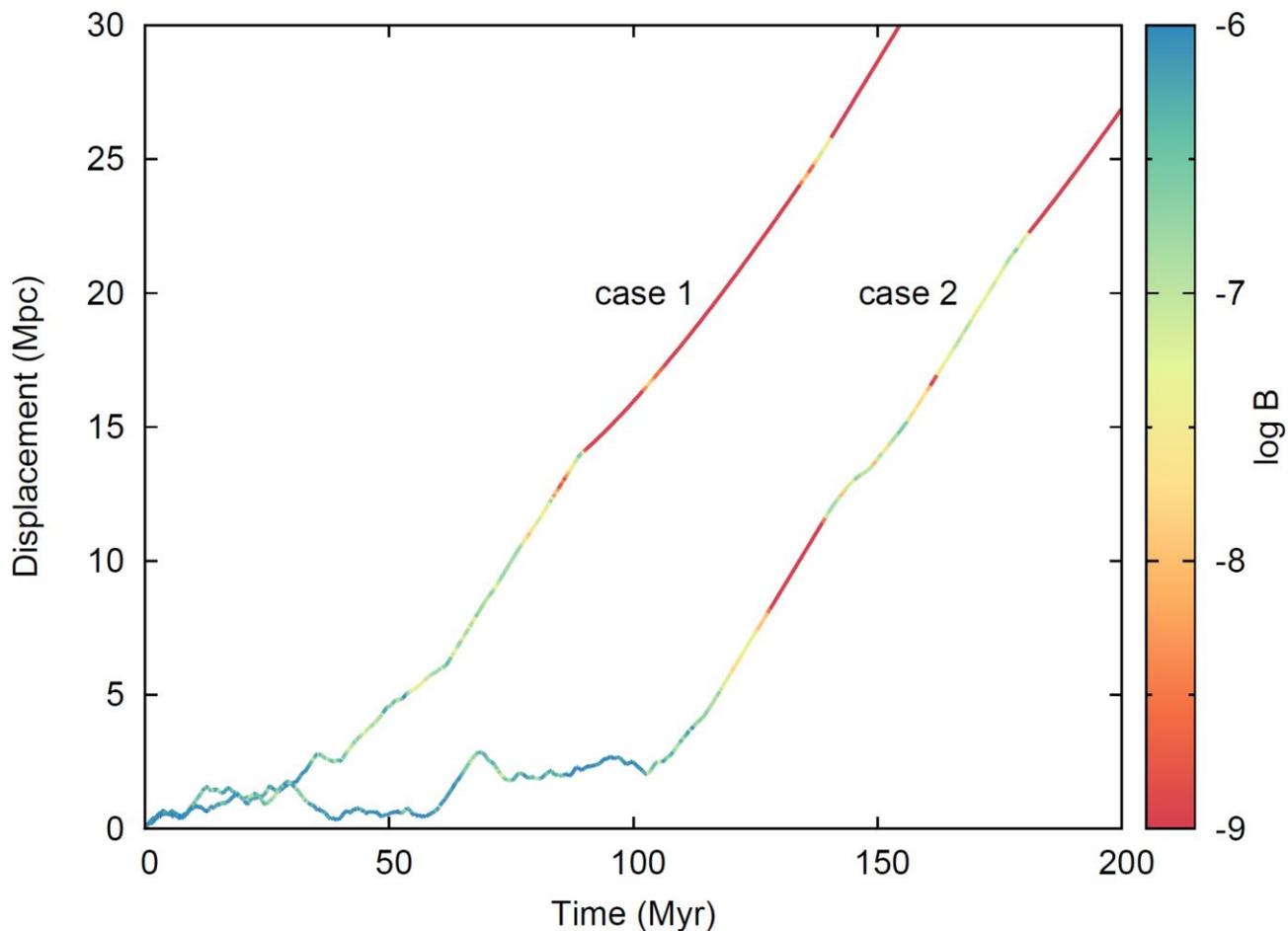

**Figure 6. Displacement of two UHE protons, shown in Figure 5, from their launching positions.** Case 1 is the one shown in the top panel of Figure 5, and case 2 is in the bottom panel. Colors map the magnetic field strength that the UHECRs experience.